\documentclass[vecphys]{svmult}

\usepackage{makeidx}         
\usepackage{graphicx}        
\usepackage{multicol}        
\usepackage[bottom]{footmisc}
\usepackage{astrobib}
\usepackage{journals}

\makeindex             

\begin{document}

\title*{Proper Motions in the Andromeda Subgroup}
\author{Andreas Brunthaler\inst{1},
Mark J. Reid\inst{2}, Heino Falcke\inst{3,4}, Christian Henkel\inst{1}\and
Karl M. Menten\inst{1}}
\authorrunning{Brunthaler et al.}
\institute{Max-Planck-Institut f\"ur Radioastronomie, Auf dem H\"ugel 69,
              53121 Bonn, Germany
\and Harvard-Smithsonian Center for Astrophysics, 60 Garden Street,
              Cambridge, MA 02138, USA
\and Department of Astrophysics, Radboud Universiteit Nijmegen, Postbus 9010, 6500 GL Nijmegen, The Netherlands
\and ASTRON, Postbus 2, 7990 AA Dwingeloo, the Netherlands
}
%
%
\maketitle

  This article presents results of VLBI observations of regions of
  H$_2$O maser activity in the Local Group galaxies M33 and IC\,10.
  Since all  position measurements were made relative to extragalactic
  background sources, the proper motions of the two galaxies could be
  measured. For M33, this provides this galaxy's three dimensional velocity, 
  showing that this galaxy is moving with a velocity of 190 $\pm$ 59
  km~s$^{-1}$ relative to the Milky Way. For IC\,10, we obtain a motion
  of 215 $\pm$ 42 km~s$^{-1}$ relative to the Milky Way. These measurements 
  promise a new handle on dynamical models for the Local Group and the 
  mass and dark matter halo of Andromeda and the Milky Way.

\section{Introduction}
The problem when trying to derive the gravitational potential of the
Local Group is that usually only radial velocities are known and hence
statistical approaches have to be used. Kulessa and Lynden-Bell
introduced a maximum likelihood method which requires only the
line-of-sight velocities, but it is also based on some assumptions
(eccentricities, equipartition) \cite{KulessaLynden-Bell1992}.

Clearly, the most reliable way of deriving masses is using orbits,
which requires the knowledge of three-dimensional velocity vectors
obtained from measurements of proper motions.
However, measuring proper motions of members of the Local Group is difficult. 
In recent years, the proper motions of a number of Galactic satellites have
been measured using the HST (\citeNP{PiatekPryorBristow2006} and references 
therein).
These galaxies are all closer than 150 kpc
and show motions between 0.2 and a few milliarcseconds (mas) per year. More
distant galaxies, such as galaxies in the Andromeda subgroup at distances of
$\sim$ 800 kpc, have smaller angular motions, which are currently not
measurable with optical telescopes.

\section{Proper Motions of M33 and IC\,10}

We observed H$_2$O maser emission from two star-forming regions in the disk 
of M33 associated with the HII region complexes M33/19 and IC\,133
eight times with the NRAO\footnote{The National Radio Astronomy
Observatory is operated by Associated Universities, Inc., under a cooperative
agreement with the National Science Foundation.} Very Long Baseline Array
(VLBA)
between March 2001 and June 2005 \cite{BrunthalerReidFalcke2005}. 
We observed the usually brightest maser in IC\,10-SE with the VLBA
thirteen times between February 2001 and June 2005 \cite{BrunthalerReidFalcke2007}.

The motions of 4 components in M33/19 and
6 components in IC\,133 could be followed over all epochs. The
component identification was based on the positions and radial
velocities of the maser emission. A rectilinear motion was fit to each
maser feature in each velocity channel separately. Then, the variance 
weighted average of all motions was
calculated.  This yields an average motion of the maser components in
M33/19 of 35.5 $\pm$ 2.7 $\mu$as yr$^{-1}$ in right ascension and
$-$12.5 $\pm$ 6.3 $\mu$as yr$^{-1}$ in declination relative to the
background source J0137+312. For IC\,133 one gets an
average motion of 4.7 $\pm$ 3.2 $\mu$as yr$^{-1}$ in right ascension
and $-$14.1 $\pm$ 6.4 $\mu$as yr$^{-1}$ in declination.

The observed proper motion of a maser
region in M33 can be decomposed into the motion of the masers due to the 
internal galactic rotation in M33, the apparent motion of M33 caused by 
the rotation of the Sun around the Galactic Center, and the true proper 
motion of M33 relative to the Galaxy. Since the motion of the Sun and 
the rotation of M33 \cite{CorbelliSchneider1997} are known, one can 
calculate the true proper motion of M33:
$-101\pm35$ km~s$^{-1}$ in right ascension
and
$156\pm47$ km~s$^{-1}$ in declination, relative to the center of the Milky Way.

Finally, the systemic radial velocity of M33 is $-$179 km~s$^{-1}$. The
radial component of the rotation of the Milky Way toward M33 is $-$140
$\pm$ 9 km~s$^{-1}$.  Hence, M33 is moving with $-$39 $\pm$ 9
km~s$^{-1}$ toward the Milky Way. Combining the proper motions and radial 
velocities, gives the three dimensional
velocity vector of M33. The
total velocity of M33 relative to the Milky Way is 190 $\pm$ 59
km~s$^{-1}$.

In IC\,10 only the strongest maser component was detected in all epochs.
Rectilinear motion was fit to the data and yielded a value of 6$\pm$5
$\mu$as~yr$^{-1}$ toward the East and 23$\pm$5 $\mu$as~yr$^{-1}$ toward the
North. Once again, the contributions of the known motion of the Sun  and the 
known rotation of IC\,10 \cite{WilcotsMiller1998} can be calculated. The true 
proper motion of  IC\,10 is $-122\pm31$ km~s$^{-1}$ in right ascension 
and
$97\pm27$ km~s$^{-1}$ in declination.

The measured systematic heliocentric velocity of IC\,10 ($-344\pm3$
km~s$^{-1}$) is the sum
of the radial motion of IC\,10 toward the Sun and the component of the
solar motion about the Galactic Center toward IC\,10 which is
$-196\pm$10 km~s$^{-1}$. Hence, IC\,10 is moving with 148$\pm$10
km~s$^{-1}$ toward the Sun.
The proper motion and the radial velocity combined give the
three-dimensional space velocity of IC\,10.  The
total velocity is 215$\pm$42~km~s$^{-1}$ relative to the Milky Way.

\section{Local Group Dynamics and Mass of M31}

If IC\,10 or M33 are bound to M31, then the velocity of the two galaxies
relative to M31 must be smaller than the escape velocity and one can
deduce a lower limit on the mass of M31:

\begin{eqnarray}
M_{M31}>\frac{v_{rel}^2R}{2G}.\nonumber
\end{eqnarray}

\begin{figure}
\center{M$_\mathrm{M31}$ [M$_\odot$]}
{\includegraphics[width=9cm,bbllx=0cm,bburx=35.5cm,bblly=1.0cm,bbury=5.5cm,clip=,angle=0]{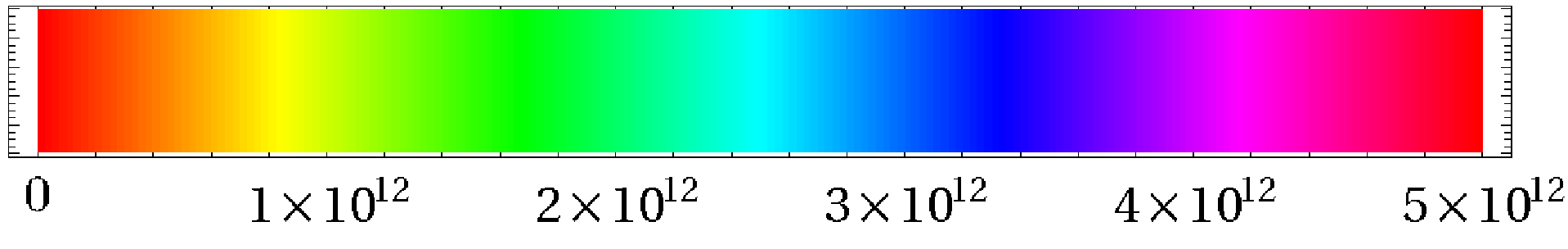}}
{\includegraphics[width=9cm,bbllx=2.5cm,bburx=13cm,bblly=16.9cm,bbury=25.5cm,clip=,angle=0]{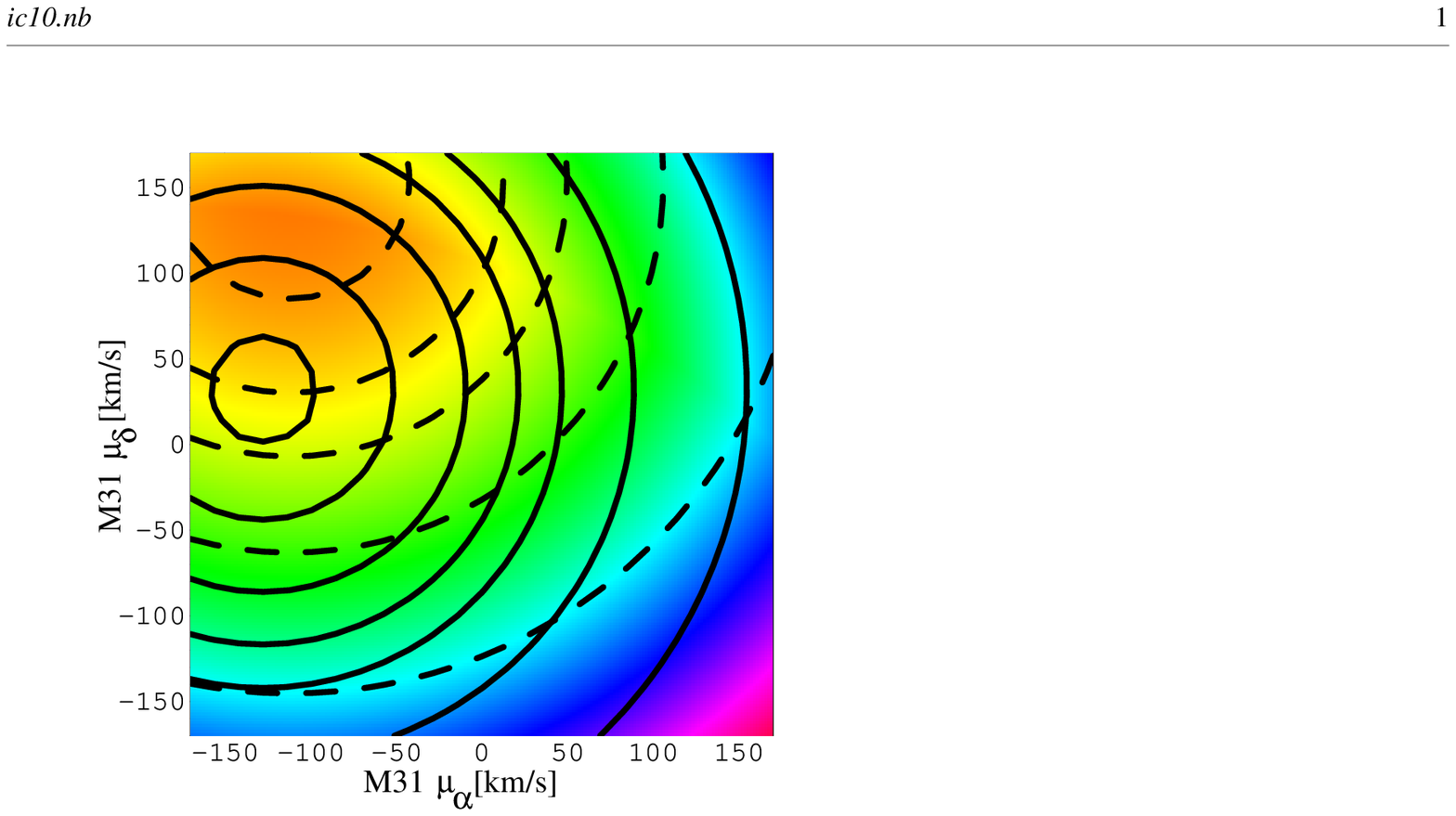}}
{\includegraphics[width=9cm,bbllx=2.5cm,bburx=13cm,bblly=16.9cm,bbury=25.5cm,clip=,angle=0]{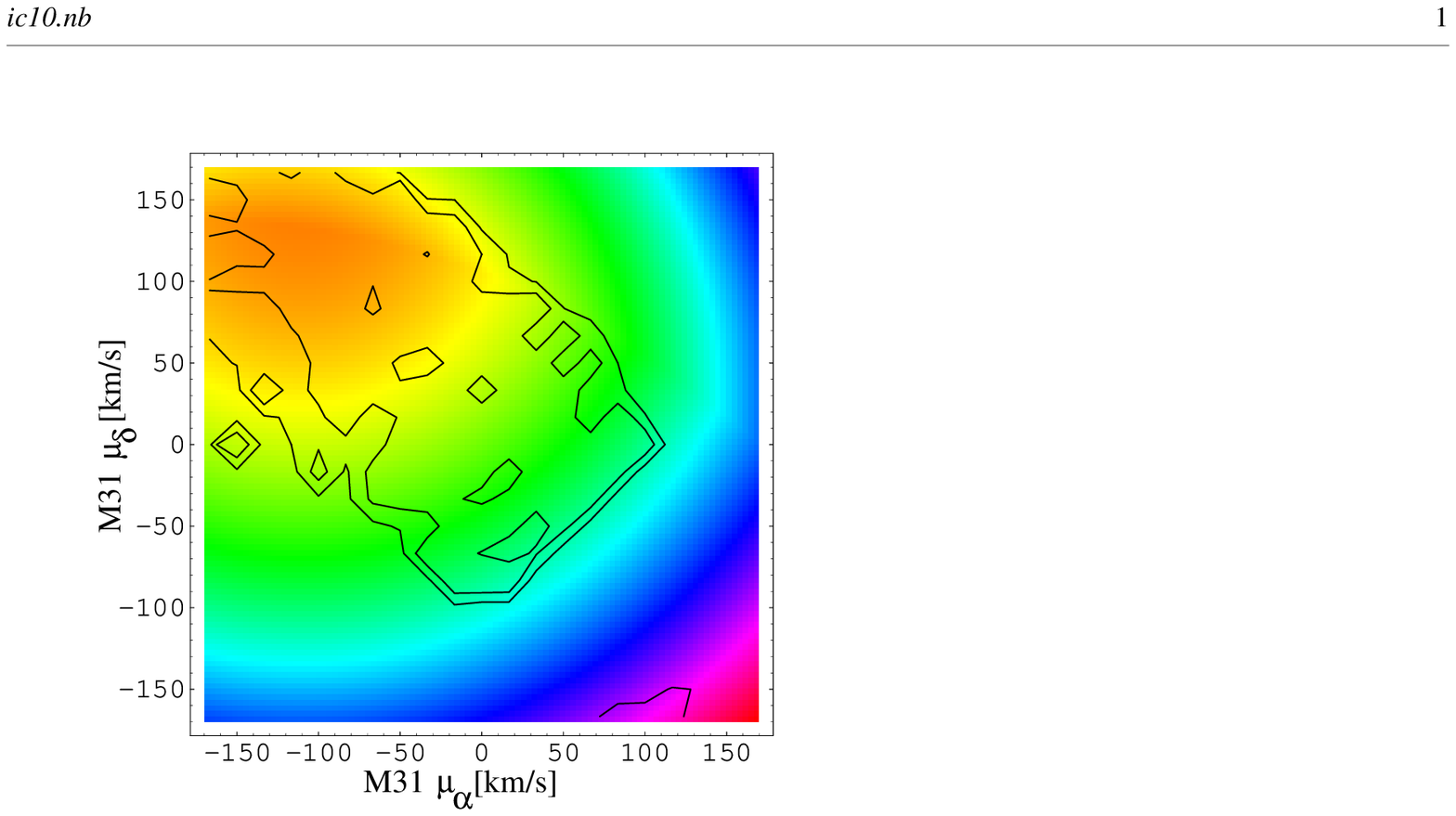}}
\caption{{\bf Top:} Lower limit on the mass of M31 for different tangential
motions of M31 assuming that M33 (dashed) or IC\,10 (solid) are bound to M31.
The lower limits to the mass of M31 are 
(4, 5, 7.5, 10, 15, 25)$\times10^{11}$M$_\odot$ for M33,
and (0.7, 1, 2.5, 5, 7.5, 10, 15, 25)$\times10^{11}$M$_\odot$ for IC\,10,
rising from inside. The colour scale indicates the maximum of both values.
{\bf Bottom:} The colour scale is the same as above and gives the
lower limit on the mass of M31. The contours show ranges of proper
motions that would have led to a large amount of stars stripped from the disk
of M33 through interactions with M31 or the Milky Way in the past.
The contours delineate 20\% and 50\%  of the total number of stars stripped
\protect\cite{LoebReidBrunthaler2005}. These regions can be excluded, since the
stellar disk of M33 shows no signs of such interactions. Taken from
\protect\cite{BrunthalerReidFalcke2007}.}
\label{mass-m31}
\end{figure}

A relative velocity of 147 km~s$^{-1}$ -- for a zero tangential motion of M31
-- and a distance of 262 kpc between IC\,10 and M31 gives a lower limit of
6.6 $\times 10^{11}$M$_\odot$. One can repeat this calculation for any
tangential motion of M31. The results are shown in Fig.~\ref{mass-m31} (top).
The lowest
value of 0.7 $\times 10^{11}$M$_\odot$ is found for a tangential motion of M31
of --130 km~s$^{-1}$ toward the East and 35 km~s$^{-1}$ toward the North.

For a relative motion of 230 km~s$^{-1}$ between M33 and M31 -- again for a
zero tangential motion of M31 -- and a distance of 202 kpc, one gets a lower
limit of 1.2 $\times 10^{12}$M$_\odot$ \cite{BrunthalerReidFalcke2005}.
Fig.~\ref{mass-m31} (top) shows also the lower limit of the mass of M31 for
different
tangential motions of M31 if M33 is bound to M31. The lowest value is
4 $\times 10^{11}$M$_\odot$ for a tangential motion of M31 of
--115 km~s$^{-1}$ toward the East and 160 km~s$^{-1}$ toward the North.

In \cite{LoebReidBrunthaler2005} it was
found that proper motions of M31 in negative right ascension and positive
declination would have lead to close interactions between M31 and M33 in the
past. These proper motions of M31 can be ruled out, since the stellar disk of
M33 does not show any signs of strong interactions.

Thus, we can rule out certain
regions in Fig.~\ref{mass-m31}. This results in a lower limit of
7.5$\times 10^{11}$M$_\odot$ for M31 and agrees with a recent
estimate of $12.3^{+18}_{-6}\times10^{11}$~M$_\odot$ derived from the
three-dimensional positions and radial velocities of its satellite
galaxies \cite{EvansWilkinson2000}.

\noindent
{\it Acknowledgements:} This research was supported by the DFG Priority Programme 1177.

%
%

\bibliography{brunthal_refs}
\bibliographystyle{aa}

%


\end{document}